\documentclass[fleqn,usenatbib]{mnras}

\usepackage[T1]{fontenc}
\usepackage{subfigure}
\usepackage{pdflscape}
\usepackage{color}
\usepackage{nccmath}
\usepackage{graphicx}
\usepackage{amsmath}
\usepackage{amssymb}
\usepackage{booktabs}
\usepackage{makecell}
\usepackage{fancyvrb}
\usepackage{enumerate}

\title[The SFHs of quiescent galaxies at $z \simeq 1$]{The connection between stellar mass, age and quenching timescale in massive quiescent galaxies at $z\simeq 1$}
\author[M. L. Hamadouche et al.]{M. L. Hamadouche$^{1}$\thanks{E-mail: mham@roe.ac.uk (MLH)},
A. C. Carnall$^{1}$, R. J. McLure$^{1}$, J. S. Dunlop$^{1}$, R. Begley$^{1}$, F. Cullen$^{1}$
\newauthor D. J. McLeod$^{1}$,  C. T. Donnan$^{1}$ and T. M. Stanton$^{1}$ \\
\\
$^{1}$SUPA\thanks{Scottish Universities Physics Alliance}, Institute for Astronomy, University of Edinburgh, Royal Observatory, Edinburgh, EH9 3HJ, UK}

\date{Accepted XXX. Received YYY; in original form ZZZ}

\pubyear{2022}

\begin{document}
\label{firstpage}
\pagerange{\pageref{firstpage}--\pageref{lastpage}}
\maketitle

\begin{abstract}
We present a spectro-photometric study of a mass-complete
sample of quiescent galaxies at $1.0 < z < 1.3$ with $\mathrm{log_{10}}(M_{\star}/\mathrm{M_{\odot}}) \geq 10.3$ drawn from the VANDELS survey, exploring
the relationship between stellar mass, age and star-formation history. Within our sample of 114 galaxies, we derive a stellar-mass vs stellar-age relation with a
slope of $1.20^{+0.28}_{-0.27}$ Gyr per decade in stellar mass. When combined with recent literature results, we find evidence that the slope
of this relation remains consistent over the redshift interval $0<z<4$. The galaxies within the VANDELS quiescent display
a wide range of star-formation histories, with a mean star-formation timescale of $1.5\pm{0.1}$ Gyr and a mean quenching timescale of $1.4\pm{0.1}$ Gyr. We also find a large scatter in the quenching timescales of the VANDELS quiescent galaxies, in agreement with previous evidence that galaxies at $z \sim 1$ cease star formation via multiple mechanisms.
We then focus on the oldest galaxies in our sample, finding that the number density of galaxies that quenched before $z = 3$ with stellar masses $\mathrm{log_{10}}(M_{\star}/\mathrm{M_{\odot}}) \geq 10.6$ is $ 1.12_{-0.72}^{+1.47} \times 10^{-5} \ \mathrm{Mpc}^{-3}$. Although uncertain, this estimate is in good agreement with the latest observational results at $3<z<4$, tentatively suggesting that neither rejuvenation nor merger events are playing a major role in the evolution of the oldest massive quiescent galaxies within the redshift interval $1<z<3$. 

\end{abstract}


\section{Introduction}\label{intro}

It is now well established that the local galaxy population is bi-modal in terms of colour, morphology and star-formation rate (SFR). The colour bi-modality was first observed using data from the Sloan Digital Sky Survey \citep[SDSS,][]{yorkSDSS}, with galaxies falling into two categories: a star-forming `blue cloud' and a quiescent `red sequence' \citep[e.g.][]{Strateva2001,Baldry2004}. In general, more-massive galaxies tend to be red spheroids with little ongoing star formation, whilst less-massive galaxies are mainly blue, star-forming discs. Over the past few decades, many studies have aimed to quantify the mechanisms responsible for producing the bi-modality in the galaxy population \citep[e.g.][]{dekelandbirnboim2006,PengQuenching, gabor_dave_morph_quenching,schawinski_gv_2014}. However, despite the wealth of ground-based and space-based data available, understanding exactly how the shutting down of star formation relates to this distinction between galaxy types remains hugely challenging.

Observations have shown that, even within the quiescent population, galaxies demonstrate a range of characteristics. For example, more-massive galaxies are known to have formed earlier in cosmic time and much more rapidly, with clear evidence for younger stellar populations in less-massive galaxies compared to their more-massive counterparts \citep[e.g.][]{cowie1996downsizing,Thomas2005, fontana_downsizing, fontanot2009,pacifici_d4000}. This phenomenon, referred to as  `downsizing', is commonly used to describe the relationship between quiescent galaxy stellar mass and age, and indicates that quenching varies as a function of stellar mass and redshift. Empirical spectral age indicators have provided strong constraints on this downsizing trend, with features such as the D\textsubscript{n}4000 index demonstrating positive correlations with stellar mass at redshift, $z \lesssim 1$ \citep[e.g.][]{bruzuald4000_1983,balogh_d4000,kauffman_dn4000, brinchmannMS, moresco_cos_chronometers10, zcosmos_moresco10, moresco_cosmic_hubble_16}. 

In addition to the bi-modality of the galaxy population, one of the most important observational results of the past few decades is the differing evolution of the star-forming and quiescent galaxy stellar mass functions (GSMF) across cosmic time, with the number density of quiescent galaxies apparently increasing by almost an order of magnitude since $z\simeq 2$ \citep[e.g.][]{k20_cimmatti,gdds,baldry2012,muzzin_uvista,davidzon_gsmf,mcleod2021_gsmf}. However, recent studies also point to a substantial population of massive quiescent galaxies out to $z > 3$ \citep[e.g.][]{Schreiber2018,valentino2020, carnall2020, Carnall2022c}. Together, these results allow us to quantify the quiescent galaxy fraction across cosmic time, an important observational constraint on galaxy evolution models \citep[e.g.,][]{Somerville_2015REVIEW}.

Another key result was the discovery that the sizes of quiescent galaxies have evolved much more rapidly than their star-forming counterparts 
since $z \sim 2$, and that quiescent galaxies follow a steeper stellar mass-size relation than star-forming galaxies at all redshifts \citep[e.g.][]{shen,trujillo_2006, ross_size_2013, correct_vdw_3dhst+candels, mowla_cosmos_dash}. The physical processes driving the size growth of quiescent galaxies are still not fully understood, although it is widely accepted that minor mergers play an important role in explaining the observed growth from $z \sim 2$ to the local Universe \citep[e.g. see][]{hopkins_mergers_simulations,trujillo_minor_mergers_2011,cimatti_dry_merger, ownsworth_mergers_2014}.

Considerable effort has been devoted to understanding which physical mechanisms are required to explain the observed differences in the properties of the star-forming and quiescent galaxy populations. Our understanding of quenching mechanisms relies heavily on simulations of galaxy formation. At $z<2$, simulations have been able to reproduce the observed bi-modality \citep[see][]{dave_2017,dave_2019,nelson_2018,simbaUVJ}. However, the situation becomes more complicated at higher redshifts, and it is much more difficult to identify the key physical drivers of quenching. The main mechanisms thought to cause quenching can be categorised into two distinct pathways: `mass' \citep[or `internal', see][]{Somerville_2015REVIEW} quenching, and `environmental' quenching. Locally, these two pathways are clearly distinguishable, suggesting that multiple mechanisms quench galaxies \citep[e.g.,][]{PengQuenching}.

Mass quenching is often thought to be associated with feedback processes such as radiative- or jet-mode active galactic nucleus (AGN) feedback \citep[e.g.][]{crotonAGN,gabor2011,Choi_2018quenchingmech}. Quenching attributed to galaxy-galaxy interactions (often referred to as `environmental' or `satellite' quenching) is thought to be the result of ram-pressure stripping, caused by satellite galaxies falling into larger dark matter halos, or virial shock-heating of the circum-galactic medium \citep[see][also referred to as `halo' quenching]{dekelandbirnboim2006}.

These mechanisms can be further categorised as `slow' and `fast' quenching pathways, respectively \citep[][]{schawinski_gv_2014,schreiber_quenching_mechs,bagpipespaper,MOSFIREBelli_2019}. Shorter quenching timescales are thought to be linked with quasar-mode AGN feedback, which is thought to be more prevalent at high redshift \citep[][]{wild_2016}. In contrast, it appears that the key process responsible for quenching at low redshift is the halting of gas accretion, taking place on much longer timescales of several Gyr \citep[e.g.][]{peng2015,trussler_quenching_mechs}.

Large spectroscopic surveys have facilitated increasingly sophisticated, statistical studies of galaxy physical properties at high redshift, with the aim of placing tighter constraints on the physical origins of quenching. The recently completed LEGA-C \citep[][]{legac_survey_paper} and VANDELS \citep[][]{vandels} surveys provide ultra-deep spectroscopy for hundreds of quiescent galaxies at $0.6 < z < 2.5$. These data sets, coupled with improved spectral energy distribution (SED) fitting methods \citep[e.g.][]{Carnall2019a,Leja2019}, have already enabled more-precise measurements of galaxy stellar masses, star-formation histories (SFHs) and stellar metallicities, unveiling significant correlations between these physical properties \citep[e.g.][]{wu_size_paper,Wu2021_LEGAC_TNG_D4000,beverage_metallicity_not_age,adam_metallicity}.

A key emerging result is the finding that the observed stellar mass vs stellar age relationship for $z\sim1$ quiescent galaxies is steeper than is predicted by the most recent generation of cosmological simulations (e.g. \citealt{Carnall2019b, Tacchella2022}). These new spectroscopic analyses build upon a corpus of earlier work aiming  to quantify these relationships, much of which was founded upon the use of elemental abundances as empirical proxies for formation and quenching timescales \citep[e.g.][]{thomas2005a,conroy2014,kriek_2019_metallicity}. Despite these advances in the field, continued, in-depth investigation into the physical properties of quiescent galaxies is still needed to build a thorough understanding of quenching and passive galaxy evolution. 

In \citet[][]{hamadouche22}, we investigated the links between stellar mass, age, size and metallicity using quiescent-galaxy samples from the LEGA-C \citep[][]{legac_survey_paper} and VANDELS \citep{vandels} spectroscopic surveys at $z\simeq 0.7$ and $z \simeq 1.1$, respectively. We examined stellar mass-age trends using the D\textsubscript{n}4000 index as a proxy for the stellar population age, finding that more-massive galaxies exhibit higher D\textsubscript{n}4000 values at both redshift ranges, consistent with prior evidence for the downsizing scenario at lower redshifts. In this work, we return to the VANDELS spectroscopic sample, building upon our previous results by employing full spectral fitting to probe the ages and SFHs of massive quiescent galaxies at $z \gtrsim 1$ in detail.

This study makes use of the fully completed VANDELS DR4 sample \citep{vandels_final}, which includes more than twice the number of quiescent galaxy spectra studied in the initial analysis of \cite{Carnall2019b}. Moreover, in this study we implement an improved physical model, along with additional metallicity constraints for the VANDELS sample from \cite{adam_metallicity}, to better constrain star-formation histories, stellar masses and formation and quenching times. Motivated by the ongoing challenges in quantifying the correlations between key quiescent galaxy physical properties, we begin by examining the relationship between stellar mass and age in our quiescent sample at $1.0 < z < 1.3$, and discuss these results in the context of downsizing. 


The structure of this paper is as follows. We introduce the VANDELS survey in Section \ref{data}, before providing details of our sample selection and spectral fitting technique using the {\scshape Bagpipes} code \citep[][]{bagpipespaper} in Section \ref{sample_selection}. We present our main results in Section \ref{results} and discuss them in Section \ref{discussion}. Finally, we present our conclusions in Section \ref{conclusions}. 
Throughout this paper, we assume a \cite{kroupaimf} initial mass function and the \cite{asplund2009} Solar abundance of $\mathrm{Z}_{\odot}=0.0142$. We assume cosmological parameters $H_{0}$ = $\mathrm{70 \ {km} \ {s^{-1}} \ {Mpc^{-1}}}$, $\mathrm{\Omega_{m}}$ = 0.3 and $\mathrm{\Omega_{\Lambda}}$ = 0.7 throughout. All magnitudes are quoted in the AB system.

\section{The VANDELS survey}\label{data}

VANDELS is a large ESO Public Spectroscopy Survey \citep{vandels, pentericcivandels, vandels_final} targeting the CDFS and UDS fields, and covering a total area of 0.2 deg$^2$. The survey data were obtained using the Visible Multi-Object Spectrograph \citep[VIMOS,][]{vimos_le_Fevre} on the ESO VLT. The final data release \citep[DR4;][]{vandels_final} provides spectra for a sample of 2087 galaxies, the vast majority of which (87 per cent) are star-forming galaxies in the redshift range $2.4 < z < 6.2$. However, in this study, we focus on the remaining 281 targets (13 per cent) selected as quiescent galaxies in the redshift range $1.0 < z <2.5$. 

\subsection{VANDELS sample selection}\label{passive_selection_criteria}

The VANDELS spectroscopic sample was originally drawn from a combination of four separate photometric catalogues. Two of these are the CANDELS GOODS South and UDS catalogues \citep{guo_candels_cats, galametz_candels_cats}, whilst the other two are custom ground-based catalogues (described in \citealt{vandels}), covering the wider VANDELS area outside of the CANDELS footprints.

The parent quiescent sample was selected from these photometric catalogues as follows. Objects were required to have \hbox{\textit{H}-band} magnitudes of $H \leq 22.5$, corresponding to stellar masses of $\mathrm{log_{10}}(M_{\star}/\mathrm{M_{\odot}}) \gtrsim 10$ over the redshift range of $1.0 \leq z\mathrm{_{spec}} \leq 1.3$ we focus on in this work (approximately 98 per cent of the full VANDELS quiescent sample has $z\mathrm{_{spec}}<1.5$), as well as \hbox{\textit{i}-band} magnitudes of $i\leq25$. To separate star-forming and quiescent galaxies, rest-frame \textit{UVJ} criteria were applied following \cite{williams09bicolour}. These criteria result in a sample of 812 galaxies, which we refer to as the VANDELS photometric parent sample.

\subsection{VANDELS spectroscopy}

Here we briefly summarise the VANDELS spectroscopic observations, while referring the reader to \cite{pentericcivandels} for a full description. From the parent sample of 812 quiescent galaxies described in the previous section, 281 were randomly assigned slits and observed as part of the VANDELS survey. Objects were observed for 20, 40 or 80 hours depending on their \textit{i}-band magnitudes. The observations were obtained using the MR grism, providing a median resolution of $R \sim 600$ across a wavelength range from $\lambda=4800 - 9800$ {\AA}.\ The VANDELS team manually measured spectroscopic redshifts, assigning redshift quality flags according to \cite{LeFevre2013}. In this paper we only use those galaxies with spectroscopic redshift flag 3 or 4, which has subsequently been shown to correspond to a $\simeq 99$ per cent probability of being correct \citep[][]{vandels_final}.

\begin{table*} 
\setlength{\tabcolsep}{6pt}
\renewcommand{\arraystretch}{1.4}
\caption{Details of the parameter ranges and priors adopted for the {\scshape Bagpipes} fitting of the VANDELS photometry and spectroscopy (see Section \ref{bagpipes_model}). Priors listed as logarithmic are uniform in log-base-ten of the parameter.}
\label{tab:bagpipes_model}
    \begin{tabular}{lcccccc}
     \midrule
\textbf{Component} & \textbf{Parameter} & \textbf{Symbol / Unit}  & \textbf{Range} & \textbf{Prior} & \textbf{Hyperparameters}\\ 
  \midrule 
Global & Redshift & $z\mathrm{_{spec}}$ & $z\mathrm{_{spec}} \pm$ 0.015 & Gaussian & $\mu = z\mathrm{_{spec}}$ $\sigma = 0.005$\\
\midrule
SFH & \makecell{Stellar mass formed \\ Metallicity \\ Falling slope \\ Rising slope \\ Peak time} & \makecell{$M_{\star}/\mathrm{M_{\odot}}$ \\ $Z_{\star}/\mathrm{Z_{\odot}}$ \\ $\alpha$ \\  $\beta$ \\ $\tau $/ Gyr } & \makecell{(1,  10$^{13}$) \\(0.2, 2.5)  \\ (0.1, 10$^3$) \\ (0.1, 10$^3$)\\ (0.1,  $t\mathrm{_{obs}}$) } & \makecell{log\\ log \\ log \\ log \\uniform} \\

\midrule
Dust & \makecell{Attenuation at 5500 {\AA} \\ Deviation from \cite{actualCalzettiLAW} slope \\ Strength of 2175 {\AA}\ bump} & \makecell{$A_{V}/$mag \\ $\delta$ \\ \textit{B} } & \makecell{ (0, 4) \\ ($-0.3, 0.3$)\\ (0, 5)} & 
\makecell{uniform \\ Gaussian \\ uniform} 
& \makecell{  \\ $\mu=0.0$, $\sigma = 0.1$\\ }\\
\midrule
Calibration & \makecell{Zeroth order \\ First order \\ Second order} &\makecell{$P_{0}$\\$P_{1}$\\$P_{2}$} & \makecell{(0.5, 1.5)\\($-0.5, 0.5$)\\($-0.5, 0.5$)} & \makecell{Gaussian \\ Gaussian \\ Gaussian} & \makecell{$\mu=1.0$, $\sigma = 0.25$ \\ $\mu=0.0$, $\sigma = 0.25$ \\ $\mu=0.0$, $\sigma = 0.25$} \\ 
\midrule
Noise & \makecell{White-noise scaling \\ Correlated noise amplitude \\ Correlation length}& \makecell{$a$ \\ $b/f_\mathrm{{max}}$ \\$l/\Delta\lambda$}& \makecell{(0.1, 10)\\(0.0001, 1)\\(0.01, 1)}& \makecell{log\\log\\log} & \\ 

\midrule \\
\end{tabular}
\end{table*}

\section{Methodology and sample selection}\label{sample_selection}

The VANDELS observations described in Section \ref{data} produce an initial sample of 269 quiescent galaxies with robust spectroscopic redshifts, of which 87 per cent have \hbox{$1 < z\mathrm{_{spec}} <  1.5$}. In this section, we describe the selection of the final quiescent sample that we use for our analysis. 

\subsection{Spectro-photometric fitting}\label{bagpipes_model}
We use {\scshape Bagpipes} \citep[][]{bagpipespaper} to simultaneously fit the available spectroscopic and photometric data for our initial sample of 269 quiescent galaxies. We incorporate several improvements to the model used to fit the VANDELS photometric catalogues in \cite{hamadouche22} (based on \citealt{Carnall2019b}), which we briefly describe below. 

We use a double-power-law star-formation history model, employing the updated 2016 versions of the BC03 stellar population synthesis models \citep[][]{bruzualcharlot2003,charlot_chevallard_2016_BC03}. We also vary the stellar metallicity from $Z_* = 0.2 - 2.5 \mathrm{Z}_{\odot}$ using a logarithmic prior. We use the \cite{Salim_2018} dust attenuation law, which parameterises the dust-curve shape through a power-law deviation, $\delta$, from the \cite{actualCalzettiLAW} law. Nebular continuum and emission lines are modelled using the {\scshape Cloudy} photoionization code \citep[][]{cloudycode}, using a method based on that of \cite{byler17}. We assume a fixed ionization parameter of $\mathrm{log_{10}}(U)= -3$. Full details of the free parameters and priors used in our fitting are provided in  \hbox{Table \ref{tab:bagpipes_model}}.

We take into account systematic uncertainties in the observed spectra of our galaxies by applying additive noise and multiplicative calibration models (e.g., \citealt{legac_survey_paper,Cappellari2017, Johnson2021}). We follow the approach outlined in Section 4 of \cite{Carnall2019b}, by fitting a second-order multiplicative Chebyshev polynomial to account for problems with flux calibration, and an additive Gaussian process model with an exponential squared kernel to model correlated additive noise between spectral pixels in our data.

\begin{figure*}
    \centering
    \includegraphics[width = \textwidth, ]{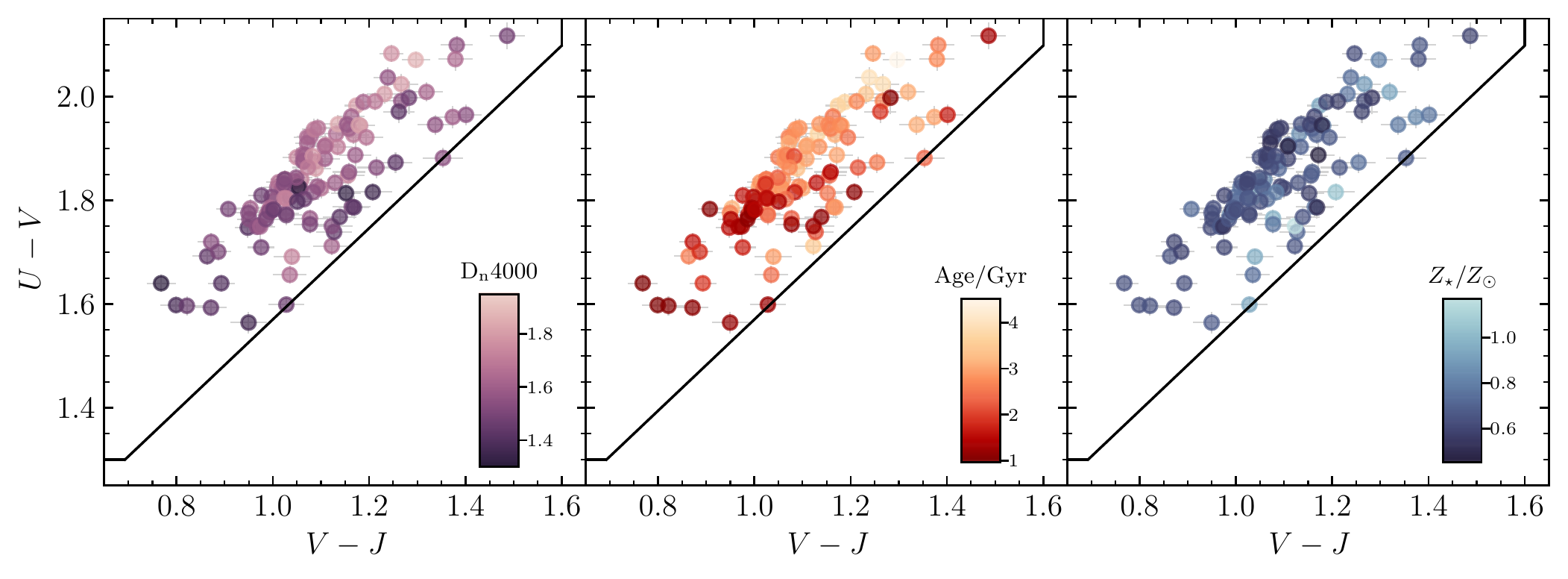}
    \caption{The distribution of the final mass-complete sample of 114 quiescent VANDELS galaxies on the \textit{UVJ} plane, highlighting trends between the rest-frame \textit{UVJ} colours and (left to right) D\textsubscript{n}4000, mass-weighted age and stellar metallicity. The first two panels show that redder rest-frame \textit{UVJ} colours correlate with higher D\textsubscript{n}4000 values and older mass-weighted ages, consistent with literature results \citep[e.g.,][]{MOSFIREBelli_2019,Carnall2019b}. The last panel, colour-coded by metallicity, does not demonstrate any significant trend.}
    \label{fig:uvj}
\end{figure*}

\subsection{A mass-complete sample}\label{mass-complete_selection_vandels}

To ensure that our final sample is mass complete, we restrict the sample to $\mathrm{log_{10}}(M_{\star}/\mathrm{M_{\odot}}) \geq 10.3$ and $1.0 \leq z_{\mathrm{spec}} \leq 1.3$ \citep[see][]{Carnall2019b}. In addition, we require members of the sample to have \hbox{$U - V > 0.88 \times (V - J) + 0.69$}, in order to remove green-valley galaxies. This has been shown to be broadly equivalent to a specific SFR cut of sSFR < 0.2/$t_\mathrm{H}$, where $t_\mathrm{H}$ is the age of the Universe at the relevant redshift \citep{bagpipespaper}. These criteria produce a sample of 139 quiescent galaxies. 

To clean the quiescent sample of potential X-ray contaminants, we remove five objects with matches in either the \textit{Chandra} Seven Mega-second catalogue \citep[][]{luo_2017} or the X-UDS catalogue \citep[][]{kocevski_2018_chandra} that cover the CDFS and UDS fields, respectively. All five galaxies with X-ray matches also display strong [O\,\textsc{ii}] emission in their rest-frame UV spectra. We also search for potential radio-loud AGN using the Very Large Array (VLA) 1.4 GHz data available for both fields \citep[][]{simpson_2006,bonzini_2013}, finding one additional AGN candidate. This object was not removed from the quiescent sample because it does not display strong [O\,\textsc{ii}] emission.

Finally, we remove one galaxy whose spectrum is highly contaminated (due to a nearby object), leaving a final, cleaned sample of 114 quiescent VANDELS galaxies. This final sample is shown on the \textit{UVJ} plane in Fig. \ref{fig:uvj}, colour-coded by mass-weighted age, D\textsubscript{n}4000 and stellar metallicity. 

\subsection{Stacked spectra}\label{stacking_analysis}
In the sections of our analysis where we make use of stacked spectra, we use the following standard procedure to produce our stacks. We first de-redshift and then re-sample each individual spectrum onto a uniform 2.5\,{\AA} wavelength grid using the spectral re-sampling module {\scshape SpectRes} \citep{spectres}. Prior to stacking, we normalise by the median flux across the wavelength range $3500 - 3700$ {\AA}. The median flux across all spectra in each pixel is then calculated. Uncertainties in the stacked spectra are calculated using the standard error on the median. 

For stacked spectra where we wish to show correlations with D\textsubscript{n}4000, the spectrum is then normalised by the median flux in the blue continuum band of the D\textsubscript{n}4000 index, such that the median flux density in the red continuum band corresponds to the D\textsubscript{n}4000 index of the stacked spectrum. We calculate D\textsubscript{n}4000 using the same prescription outlined in Section 3.4 of \cite{hamadouche22}.

\subsection{Size measurements}
We use the {\scshape Galfit} \citep{galfit_paper} size measurements from \cite{hamadouche22} for 110/114 galaxies in the final sample. For the remaining galaxies we adopted an identical procedure to \cite{hamadouche22}, using \textit{HST} F160W images for the three galaxies within the CANDELS footprint and \textit{HST} ACS F850LP imaging in CDFS for the single galaxy lying outside the CANDELS footprint. 

\begin{figure*}
\includegraphics[width=\linewidth]{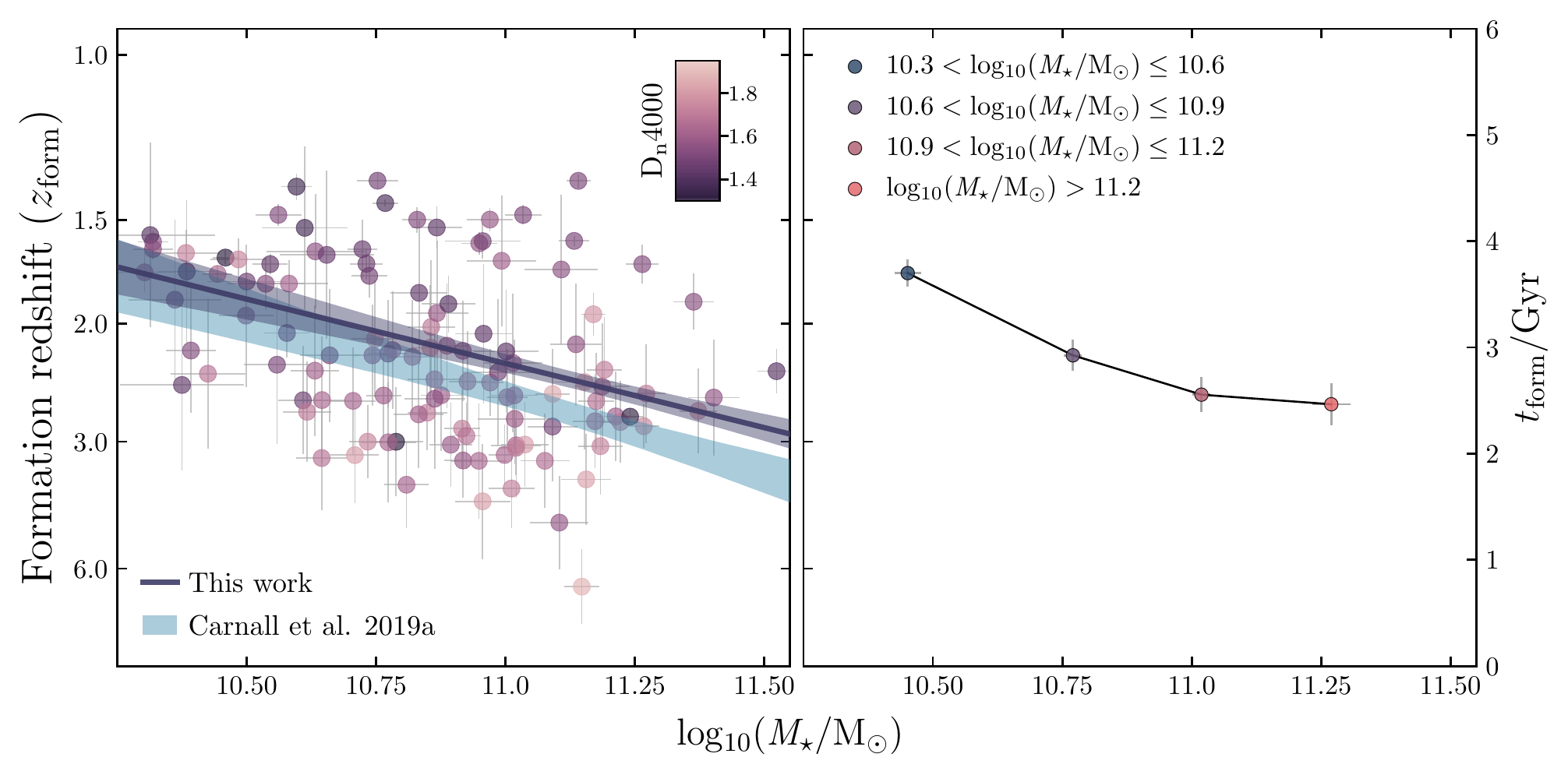}
\caption{\textit{Left:} Stellar mass versus formation redshift for our final mass-complete VANDELS DR4 quiescent galaxy sample. The sample is colour-coded by D\textsubscript{n}4000, demonstrating a clear preference for higher D\textsubscript{n}4000 values at earlier formation times. The relationship we fit in Section \ref{results:agemass} is shown in purple, with the  $1 \sigma$ confidence interval shaded. The relation derived for a smaller sample from VANDELS DR2 by \protect\cite{Carnall2019b} is shown in blue. \textit{Right:} Median formation redshifts for our sample in 0.3-dex bins. The error bars are the standard errors for $t \mathrm{_{form} / \ Gyr}$ and $\mathrm{log}_{10} (M_{\star} / \mathrm{M_{\odot}})$ in each stellar-mass bin. A clear negative correlation is observed. The final bin contains only ten objects above $\mathrm{log}_{10}(M_{\star} / \mathrm{M_{\odot}}) > 11.2 $, meaning that the apparent flattening of the relationship is challenging to assess.}
\label{fig:tfvmass} 
\end{figure*}

\section{Results}\label{results}
In this section, we present the results obtained from full-spectral fitting of our final quiescent galaxy sample.

\subsection{Trends with rest-frame \textit{UVJ} colour} 

In Fig. \ref{fig:uvj}, we show the distribution of the final mass-complete sample of VANDELS quiescent galaxies (see Section \ref{mass-complete_selection_vandels}) on the rest-frame \textit{UVJ} diagram, coloured by mass-weighted age, D\textsubscript{n}4000, and stellar metallicity. In the first panel, we see that the galaxies with redder \textit{U--V} and \textit{V--J} colours, also tend to have higher mass-weighted ages, consistent with recent literature results \citep[e.g.,][]{MOSFIREBelli_2019, Carnall2019b}. In the next panel, the trend with D\textsubscript{n}4000 is similar; lighter-coloured points indicate higher D\textsubscript{n}4000 values, which is consistent with the trend seen in other samples at similar redshifts \citep[e.g.][]{whitaker2013}.
The final panel of Fig. \ref{fig:uvj} shows the sample  coloured by metallicity. There is no obvious trend between metallicity and \textit{UVJ} colour. The individual metallicities we measure are however consistent with scattering around the median value of log$_{10}(Z_*/$Z$_\odot) = -0.13\pm0.08$ determined from an optical+NIR stack at  $z \sim 1.15$ by \cite{adam_metallicity}.

\subsection{The relationship between stellar mass and age}\label{results:agemass}
We present our results for the stellar-mass vs age relation in Fig. \ref{fig:tfvmass}. 
We plot redshift of formation, $z\mathrm{_{form}}$, against stellar mass. The right-hand axis shows the corresponding formation time, $t\mathrm{_{form}}$, measured forwards from the Big Bang. In this paper, we take $t\mathrm{_{form}}$ and $z\mathrm{_{form}}$ to be the age of the Universe and redshift corresponding to the mass-weighted age of the galaxy. We see a clear negative correlation, albeit with considerable scatter. A trend is also visible between $t_\mathrm{form}$ and D\textsubscript{n}4000 in Fig. \ref{fig:tfvmass}, with galaxies that have earlier formation times exhibiting higher values of D\textsubscript{n}4000, as would be expected. 

We fit a linear relationship between $t_\mathrm{form}$ and log$_{10}(M_*/$M$_\odot)$, including an intrinsic scatter term, using the nested sampling Monte Carlo algorithm MLFriends \citep[][]{buchner16,buchner19} using the UltraNest\footnote{\url{https://johannesbuchner.github.io/UltraNest/}} package \citep[][]{ultranest_buchner}. We derive a best-fitting relation of:

\begin{equation}\label{eq:mass_age_eq}
    (t\mathrm{_{form} \ / \ Gyr}) = 2.85^{+ 0.08}_{-0.09} - 1.20^{+0.28}_{-0.27} \ \mathrm{log_{10}}(M_{\star}/ 10^{11} \ \mathrm{M_{\odot}}).
\end{equation}

\begin{figure*}
    \centering
    \includegraphics[width = \linewidth]{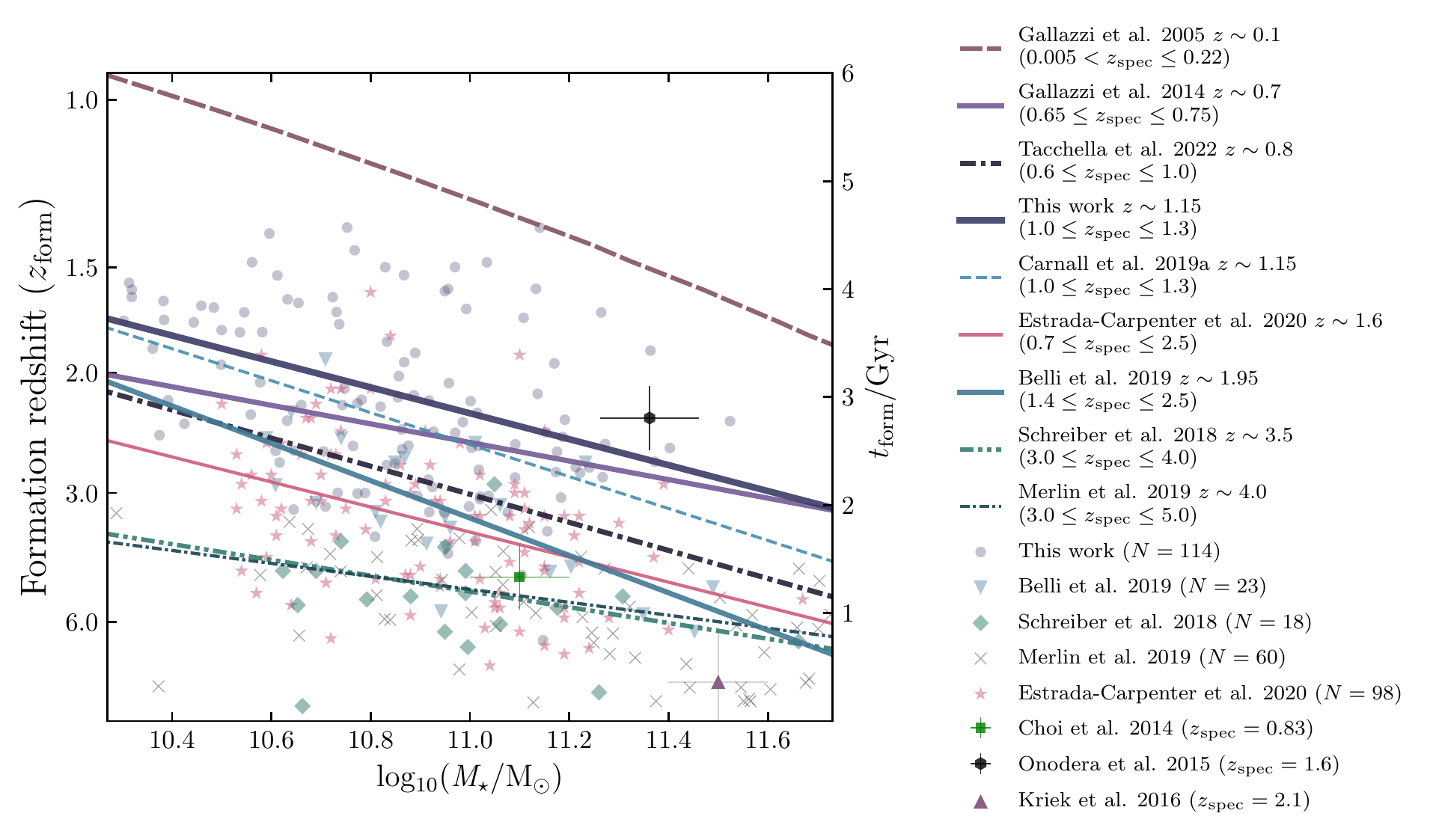}
    \caption{Stellar mass versus formation redshift for massive quiescent galaxies, taken from a range of studies across a wide range in observed redshift. Our sample of VANDELS DR4 quiescent galaxies are shown as circles, and our best-fit line is over-plotted. For some studies that do not report best-fitting relationships between stellar mass and age, we have fitted their results for individual galaxies using the methodology described in Section \ref{results:agemass}. These studies broadly agree on the slope of the relationship, which is found to be consistent at $\simeq1.5$ Gyr per decade in mass across cosmic history. The normalisation of these relationships does not follow the expected smooth evolution with observed redshift, likely due to methodological differences (see Section \ref{discussion:agemass:norm}).}
    \label{fig:literatureplot}
\end{figure*}

We also find an intrinsic scatter of $(t\mathrm{_{form} \ / \ Gyr}) = 0.51^{+ 0.09}_{-0.07}$. 
We show the fit to our data in Fig. \ref{fig:tfvmass} (purple line) with the shaded region showing the 1$\sigma$ confidence interval. 

We also show the result derived by \cite{Carnall2019b}, using the VANDELS DR2 sample of 53 galaxies, which is a subset of our new 114-galaxy final VANDELS sample. The slope of our new relation is in good agreement with this previous result, however we recover a $\sim300$ Myr offset towards younger ages. 

To explore the origin of this offset, we re-fit our linear model to the sub-sample of 53 galaxies used by \cite{Carnall2019b}, obtaining a result consistent with theirs. We therefore conclude that this offset is a result of our expanded statistical VANDELS DR4 sample, which contains more galaxies that have high stellar masses and lower formation redshifts with respect to the DR2 subset.

We also explore the median relationship between stellar mass and age in our sample by binning our galaxies into equal-width stellar-mass bins of 0.3 dex. This is shown in the right-hand panel of Fig. \ref{fig:tfvmass}, where the relationship is clear up to stellar masses of $\mathrm{log_{10}}(M_{\star}/\mathrm{M_{\odot}}) \simeq 11.2$. We discuss this relationship in more detail in Section \ref{formationtimes}, making comparisons to relevant literature, which are shown in Fig. \ref{fig:literatureplot}.

\subsection{The oldest galaxy at $z\simeq 1$}\label{oldestgals}

From inspection of Fig. \ref{fig:tfvmass}, it is clear that there is a single galaxy (ID: 111129) which falls significantly below the main stellar mass vs age distribution, with a formation time of $t\mathrm{_{form}} = 0.75^{+0.41}_{-0.29}$ Gyr ($z_{\rm form}=7.02^{+3.06}_{-2.07}$), and a quenching time ($t\mathrm{_{quench}}$ is defined as the age of
the Universe at which the normalised star-formation rate, nSFR, as defined in
\citealt{bagpipespaper}, first falls below 0.1) of $t_{\mathrm{quench}}$ = $1.94 ^{+0.86}_{-0.67}$ Gyr ($z_{\rm quench}~=~3.23^{+1.41}_{-0.93}$) after the Big Bang, respectively. Given recent reports, based on the first data from JWST, of the assembly of significant numbers of massive galaxies during the first billion years (e.g., \citealt{Labbe2022}), and their subsequent quenching during the second billion years (e.g., \citealt{Carnall2022c}), this is clearly an object of significant interest.



\begin{figure}
    \centering
    \includegraphics[width = \columnwidth]{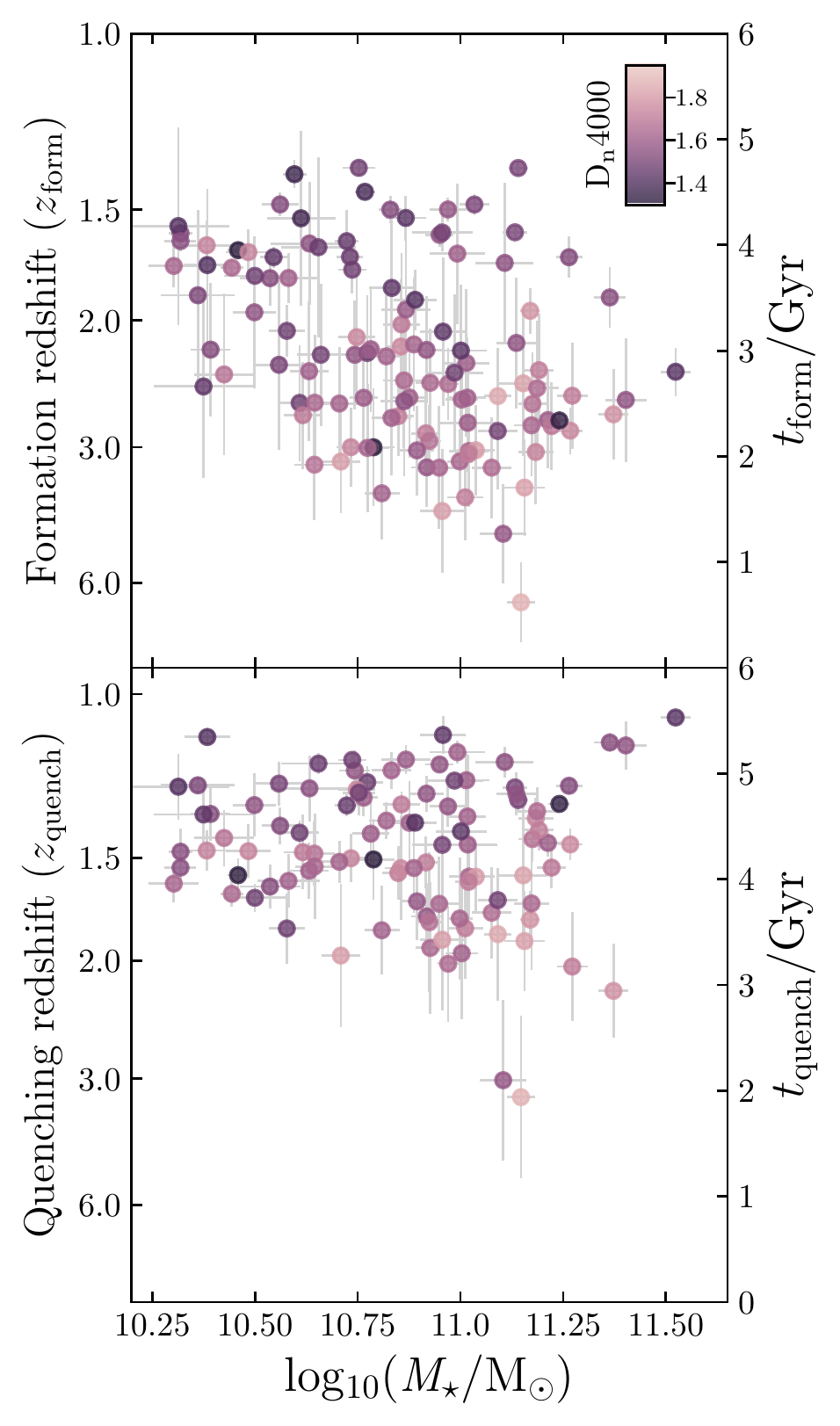}
    \caption{Stellar mass versus formation redshift and quenching redshift (in the top and bottom panels respectively) for our final sample of VANDELS quiescent galaxies. The galaxies are colour-coded by D\textsubscript{n}4000, highlighting that higher values of D\textsubscript{n}4000 are observed for galaxies with earlier formation times.}
    \label{fig:tqvmass}
\end{figure}

\section{Discussion} \label{discussion}

In Section \ref{results}, we report the relationship between stellar mass and age from full spectral fitting of our mass-complete VANDELS quiescent sample. In this section, we discuss our results, focusing on relationships between age, \textit{UVJ} position,  D\textsubscript{n}4000 and metallicity evident within our sample. 

\subsection{The formation times of quiescent galaxies}\label{formationtimes}

Over the past decade, extensive research has been conducted into the star-formation histories and ages of quiescent galaxies. This has revealed a sub-population of extremely old galaxies, which formed very early in cosmic history (e.g., \citealt{glazebrook17, Schreiber2018, valentino2020}). These galaxies tend to have higher stellar masses and more compact morphologies than is typical for the quiescent population. In order to constrain the build-up of the quiescent population across cosmic time, and reveal the fate of these oldest, most extreme systems, detailed knowledge of the stellar mass vs stellar age relationship as a function of observed redshift is required. The stellar mass vs stellar age relation presented in Section \ref{results:agemass} is based on the robust, mass-complete VANDELS spectroscopic sample. In this section, we compare these results with similar studies in the literature, across a broad redshift range.

Our results are placed into the context of recent literature in Fig. \ref{fig:literatureplot}, which shows results derived by \cite{Gallazzi2005,Gallazzi2014,choi_2014_mzr,onodera2015,Schreiber2018,MOSFIREBelli_2019,Carnall2019b,merlin_quenching,estradacarpenter2020} and \cite{Tacchella2022}, with the stellar mass vs age relations derived in various observed redshift ranges over-plotted. For several data-sets shown in the figure, no average relationship between stellar mass and age is calculated by the authors. In these cases, we perform a fit to the individual galaxy masses and ages, using the same method outlined in Section \ref{results:agemass}.

\subsubsection{Slopes of the observed relationships}

The slope of the stellar mass vs age relationship is intimately connected to the physics of quenching in massive galaxies. We derive a slope for the VANDELS DR4 sample of 1.20$_{-0.27}^{+0.28}$ Gyr per decade in mass. As can be seen from Fig. \ref{fig:literatureplot}, this is in good agreement with the other literature relationships shown. At the highest redshifts, the sample of \cite{Schreiber2018} at $3.0 < z < 4.0$ displays a slope consistent with our result at $z \simeq 1.1$ to within 2$\sigma$. In the local Universe, the results of \cite{Gallazzi2005} also display a very similar slope. This suggests the slope of the stellar mass vs age relationship for massive quiescent galaxies remains broadly constant across cosmic history.

As can be seen from Fig. \ref{fig:literatureplot}, the results of \cite{MOSFIREBelli_2019}, who study a sample of 23 massive quiescent galaxies at $1.5 < z < 2.5$ using data from the Keck-MOSFIRE spectrograph, suggest a steeper relation between stellar mass and age. We perform a fit to their galaxies on the stellar mass-age plane, finding a slope of 1.73$^{+0.40}_{-0.40}$ Gyr per decade in mass. Whilst this is a steeper slope than our result, it is not strongly in tension, owing to the relatively small samples involved.

In \cite{Carnall2019b} and \cite{Tacchella2022}, the authors compare the observed stellar mass vs age relationship with the predictions of cosmological simulations. \cite{Carnall2019b} derive this relationship from snapshots of the 100 $h^{-1}$ Mpc box runs of {\scshape Simba} \citep[][]{dave_2019} and {\scshape IllustrisTNG} \citep[][]{nelson_2018} at $z = 0.1$ and $z = 1.0$. They find that these simulations predict slopes of $\simeq$1.5 Gyr per decade in mass in the local Universe, but much shallower slopes at $z\sim1$, with \cite{Tacchella2022} reporting similar findings for IllustrisTNG at $z\sim0.7$.

Our results are consistent with the predicted slopes of $\sim$1.5 Gyr per decade in mass from these two simulations in the local Universe ($z \sim 0.1$). However, our results again suggest that simulations should seek to reproduce the same, steeper stellar mass vs age relationship for massive quiescent galaxies throughout cosmic history.

\subsubsection{Normalisations of the observed relationships}\label{discussion:agemass:norm}

The redshift evolution of the average age of quiescent galaxies at fixed stellar mass is influenced primarily by the quenching of new galaxies that join the quiescent population over time (e.g. \citealt{mcleod2021_gsmf}). This effect is sometimes known as progenitor bias. The expected evolution of the relationships shown in Fig. \ref{fig:literatureplot} due to progenitor bias would be a steady increase in normalisation from high to low redshift. The rate of this decrease in formation redshift with decreasing observed redshift is also highly sensitive to the effects of merger and rejuvenation events.

Unfortunately, this idealised smooth upward evolution with decreasing redshift is not observed in Fig. \ref{fig:literatureplot}. Whilst studies of galaxy samples at the highest observed redshifts typically return the highest formation redshifts, and the local-Universe study of \cite{Gallazzi2005} returns the lowest, there is confusion between these extremes. We report later average formation times than all studies targeting higher-redshift galaxy samples; however, both \cite{Gallazzi2014} and \cite{Tacchella2022} also report earlier average formation times than our study, despite analysing samples at lower observed redshifts ($z\simeq0.7$, as opposed to $z\simeq1.1$ for our sample).

As has been discussed in several recent works \citep{Tacchella2022,adam_metallicity}, these differences are likely the result of methodological differences between studies. The two most important of these are different definitions of age (mass-weighted vs light-weighted), and differences between parametric and non-parametric SFH models, the latter of which typically return older stellar ages \citep{Carnall2019a,Leja2019,Leja2019b}. For this reason, a clear understanding of the redshift evolution of the normalisation of this relationship requires a study applying the same methodology to observed samples at a wide range of redshifts.

\begin{figure*}
\centering
\includegraphics[width =\linewidth]{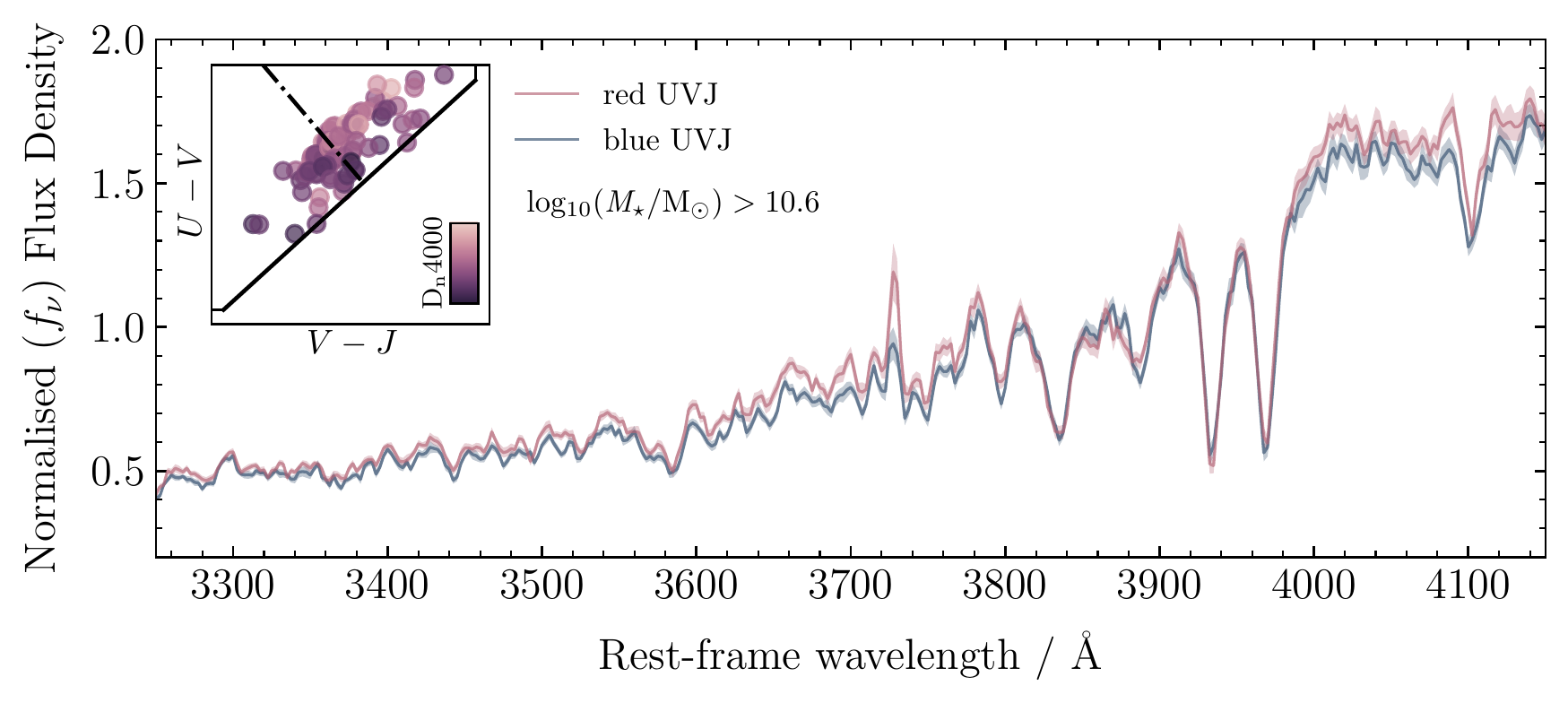}
\caption{The stacked spectra of our final sample of galaxies (for galaxies with $\mathrm{log_{10}}(M_{\star}/\mathrm{M_{\odot}}) \geq 10.6$) in two bins based on their position on the \textit{UVJ}-diagram. The galaxies are split into blue \textit{UVJ} and red \textit{UVJ} populations (see inset) following a method similar to the one adopted by \citet{whitaker2013}. As detailed in the main text, we see an increase in D\textsubscript{n}4000 and median mass-weighted age with increasing \textit{UVJ} colour (from blue to red).}
\label{fig:uvjstack}
\end{figure*}

\subsubsection{Number density of the oldest galaxies at $z\simeq 1$}

Fig. \ref{fig:tqvmass} shows the formation and quenching redshifts of our VANDELS quiescent sample versus stellar mass, in the top and bottom panels, respectively. A significant number of galaxies have formation redshifts of $z_\mathrm{form}>3$, all of which have stellar masses of $\mathrm{log_{10}}(M_{\star}/\mathrm{M_{\odot}}) \geq 10.6$. Only one galaxy has a formation redshift of $z\mathrm{_{form}} > 5$ (see Section \ref{oldestgals}).

Only two galaxies in our sample have quenching redshifts $z_\mathrm{quench}$ > 3. These objects are of particular interest, given that current simulations seem to under-predict the numbers of galaxies that quenched at these very early times \citep[see][]{Schreiber2018,Cecchi2019,Tacchella2022}, possibly due to an additional mechanism capable of causing a rapid early shutdown of star formation, not yet included in simulations. We calculate the number density of galaxies in our VANDELS sample that have $z_\mathrm{quench} > 3$, recovering a value of $1.12_{-0.72}^{+1.47} \times 10^{-5} \ \mathrm{Mpc}^{-3}$ \citep[with Poisson uncertainties calculated using the confidence intervals presented in][]{gehrels1986}. This is consistent with the results of \cite{Schreiber2018}, who calculate a number density for quiescent galaxies observed at $3 < z < 4$ of $(1.4 \pm 0.3) \times 10^{-5} \ \mathrm{Mpc}^{-3}$. This preliminary agreement is encouraging, though our sample of such old quiescent galaxies is very small. Our result is consistent with neither rejuvenation or mergers having a significant impact on this population from $1 < z < 3$. For example, if we had found no quiescent galaxies with $z_\mathrm{quench} > 3$ in our sample, this would suggest that the majority of $z>3$ quiescent galaxies experience mergers and/or rejuvenation by $z\simeq1$. However, much larger samples will be necessary to conduct detailed comparisons of this nature.


\begin{figure}
\centering
\includegraphics[width =\columnwidth]{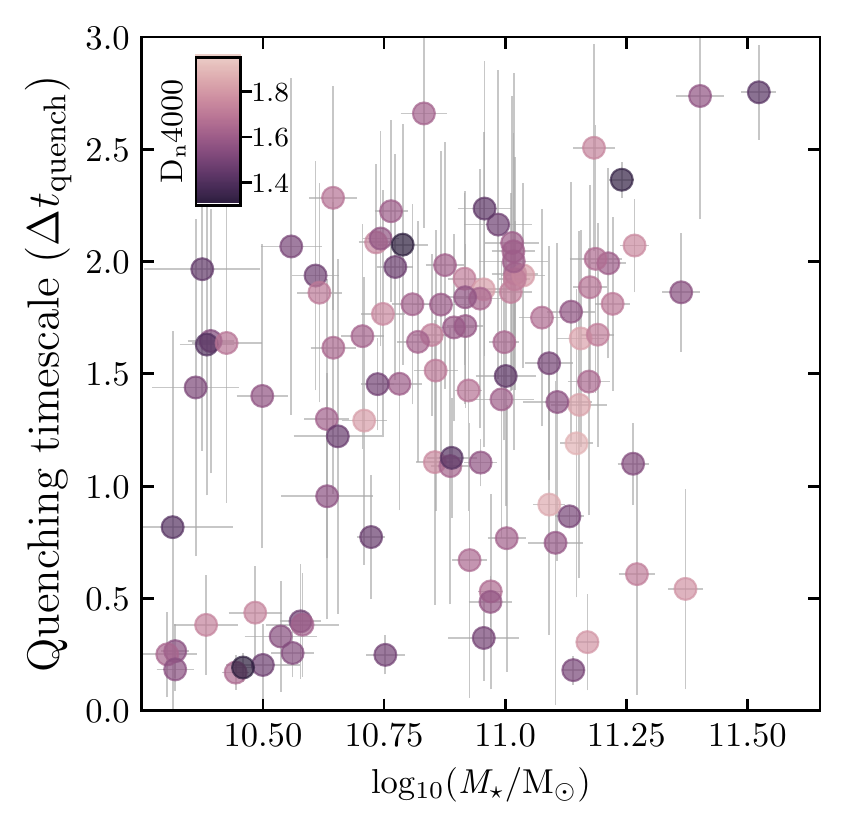}
\caption{Quenching timescale ($\Delta t_{\rm quench}$) versus stellar mass for the full VANDELS quiescent galaxy sample. The value of the D$_{\rm n}$4000 index is shown by the colour bar.}
\label{fig:new_qt}
\end{figure}

\subsection{The relationship between colour and age}
Next we consider the relationship between galaxy stellar age and position on the \textit{UVJ} diagram, performing a stacking analysis of our sample using two bins in rest-frame colour. Following the approach of \cite{whitaker2013}, we divide the sample into two bins on the \textit{UVJ} diagram, separated using the criteria:
\begin{ceqn}
\begin{align}\label{eq:uvj}
    (U-V) = -1.14 \times (V-J) + 3.10,
\end{align}
\end{ceqn}
as illustrated in the inset panel of Fig. \ref{fig:uvjstack} by the dot-dashed line. In order to minimise the impact of the correlation between stellar mass and age, we additionally restrict both \textit{UVJ} bins to only include galaxies with $\mathrm{log_{10}}(M_\mathrm{{\star}}/ \mathrm{M_{\odot}})~\geq~10.6 $. Adopting these criteria produces bins containing similar numbers of objects and comparable median stellar masses;  $\mathrm{log_{10}}(M_\mathrm{{med}}/ \mathrm{M_{\odot}})  = 10.86 \pm 0.03 $ and $11.00 \pm 0.04$ for the blue and red \textit{UVJ} bins, respectively.

\begin{table}
\setlength{\tabcolsep}{6pt}
\renewcommand{\arraystretch}{1.4}
\caption{The D\textsubscript{n}4000 values calculated from the stacked spectra shown in Fig. \ref{fig:uvjstack}. We also report median values in each of the three bins for D\textsubscript{n}4000 and mass-weighted age, where \textit{N} represents the number of galaxies in each bin. }
\label{tab:uvj_table}
    \begin{tabular}{cccccc}
     \midrule
\textit{UVJ} position & \textit{N} & D\textsubscript{n}4000\textsubscript{stack} & D\textsubscript{n}4000\textsubscript{med} & age / Gyr \\ 
 \midrule 
Blue \textit{UVJ} & 45 & $1.57\pm0.01$ & $1.58\pm0.03$ & $2.45\pm0.14$  \\ 

Red \textit{UVJ} & 47 & $1.66\pm0.02$ & $1.63\pm0.04$ & $2.80\pm0.12$  \\
\midrule 
\end{tabular}
\end{table}

In the main panel of Fig. \ref{fig:uvjstack} we show stacked spectra constructed from the objects in each \textit{UVJ} bin and in Table 2 we present D\textsubscript{n}4000 values calculated from 
the stacked spectra, together with the median D\textsubscript{n}4000 values and stellar ages of the objects in each bin. It is clear from these results that galaxies within the red \textit{UVJ} bin display larger D\textsubscript{n}4000 values and older stellar ages than their counterparts within the blue \textit{UVJ} bin. This is consistent with the expected correlation between age and \textit{UVJ} colour \citep[e.g.,][]{whitaker2013, MOSFIREBelli_2019}, and with the trends observed in the centre panel \textit{UVJ} diagram in Fig. \ref{fig:uvj}, which is colour-coded by mass-weighted age from the {\scshape Bagpipes} fits. This stacking experiment independently confirms and quantifies the age-colour trend, given the offset in median D\textsubscript{n}4000 values calculated for the red and blue \textit{UVJ} colour bins.  We note that \cite{whitaker2013} find ages of 0.9$^{+0.2}_{-0.1}$ Gyr and 1.6$^{+0.5}_{-0.4}$ Gyr for their blue and red \textit{UVJ} sub-samples, based on stacked grism spectra of quiescent galaxies at $1.4<z<2.2$. Within the large uncertainties, this difference of $0.7\pm{0.5}$ Gyr is fully consistent with the difference of $0.4\pm{0.2}$ Gyr we find from our analysis. 
\subsection{Quenching timescales}
As discussed in the introduction, recent studies of the star-formation histories
of quiescent galaxies point to the existence of multiple quenching channels \citep[e.g.,][]{MOSFIREBelli_2019,Carnall2019b,Tacchella2022}.
In general, the star-formation histories we derive for the
VANDELS quiescent sample are consistent with this picture, displaying a range of formation and quenching times.
However, to investigate this issue in more detail it is interesting to define a quenching timescale parameter: $\Delta t_\mathrm{{quench}}$. 
In this work, $t\mathrm{_{quench}}$ is defined as the age of
the Universe at which the normalised star-formation rate (nSFR; see
\citealt{bagpipespaper}) falls below 0.1, corresponding to the time
after the Big Bang at which a galaxy is labelled as quiescent by our
selection criteria. Therefore, the quenching timescale is naturally defined as:
\begin{equation}
    \Delta t_\mathrm{{quench}} = t(z\mathrm{_{quench}}) -
    t(z\mathrm{_{form}}).
\end{equation}

In Fig. \ref{fig:new_qt}, we plot quenching timescale versus stellar mass for the VANDELS quiescent galaxies.
It is clear that within our sample there is not a significant correlation between quenching timescale and stellar mass, with some of the highest mass galaxies having quenching timescales of $2-3$ Gyr, while others quench in significantly less than 1 Gyr. The mean quenching timescale for our full sample is $\Delta t_\mathrm{{quench}}~=~1.4~\pm~0.1$~Gyr. 

Although the large scatter observed within our VANDELS sample in Fig. \ref{fig:new_qt} may be a result of intrinsic galaxy-to-galaxy variations, it is still important to note the fact that properties such as star-formation histories, ages and quenching times can be affected by the fitting method \citep[see e.g.][]{pforr2012, Carnall2019a, Leja2019}.

Throughout this paper we have used a double-power law parametric SFH, which is more flexible than other parametric SFHs (due to it allowing independent rising and falling phases) and is also a better estimator of stellar masses and ages \citep{bagpipespaper}. Although parametric SFHs (such as the double-power law) appear to describe the majority of simulated galaxy populations well, they are not without fault; for example, parametric SFHs fail to model sharp transitions in SFR as accurately as non-parametric SFHs \citep{Leja2019}.

For non-parametric models, SFR and ages have been shown to be more dependent on the priors used, than e.g. stellar masses \citep{Leja2019}, however, the errors on these parameters are larger and thus more realistic than the smaller errors produced by parametric SFHs. In Fig. \ref{fig:new_qt}, the quenching timescales of our galaxies appear to be well-constrained, however these error bars may not be truly representative of the uncertainty on this parameter due to these SFH modelling effects. In future work, we will extend our methodology to extract quenching timescales from quiescent galaxy samples using non-parametric SFHs, in order to explore the effects of different approaches to  modelling star-formation histories. This will allow for a more quantitative understanding of the mechanisms by which these galaxies have quenched, and give further insight into their evolution since quenching.

\section{Conclusions}\label{conclusions}
In this paper, we have explored the relationships between stellar
mass, age, star-formation history and quenching timescales for a robust spectroscopic sample of quiescent galaxies at $1.0 < z < 1.3$.
Our main results and conclusions can be summarised as follows:

\begin{enumerate}

\item{We derive significantly improved constraints on the relationship between stellar population age and stellar mass for quiescent galaxies at $z\simeq 1.1$. From our full VANDELS sample we derive an age-mass relation which has a slope of $1.20^{+0.28}_{-0.27}$ Gyr per decade in stellar mass.}

\item{Comparing to previous studies in the literature, we find good agreement on the slope of the age-mass relation for quiescent galaxies from the local Universe out to $z\simeq 4$. The observed slope is in good agreement with the prediction from simulations at $z\simeq 0$, but significantly steeper than simulations predict at $z\geq 1$.}
  
\item{The results of our spectro-photometric fitting predict that the number density of already quenched galaxies at $z\geq3$ with
stellar masses $\mathrm{log_{10}}(M_{\star}/\mathrm{M_{\odot}}) \geq 10.6$ is $1.12_{-0.72}^{+1.47} \times 10^{-5} \ \mathrm{Mpc}^{-3}$. Although subject to large uncertainties due to small-number statistics, this estimate is in good agreement with the latest measurements at $3<z<4$. The implication is that rejuvenation or merger events are not playing a major role in modulating the number density of the oldest massive quiescent galaxies within the redshift interval $1<z<3$, although they cannot be ruled out entirely.}

\item{We confirm previously reported results that quiescent galaxies with redder \textit{UVJ} colours are systematically older than their bluer counterparts, finding an off-set of $0.4\pm{0.2}$ Gyr in the median age of mass-matched samples.}

\item{The VANDELS sample of $z\simeq 1.1$ quiescent galaxies displays a wide range of formation and quenching redshifts.
We find that the mean quenching timescale is $1.4\pm{0.1}$ Gyr, where $\Delta t_\mathrm{{quench}} = t(z\mathrm{_{quench}}) - t(z\mathrm{_{form}})$.
The oldest galaxy within the VANDELS sample (ID: 111129) has $z_{\rm form}=7.02^{+3.06}_{-2.07}$ and $z_{\rm quench}~=~3.23^{+1.41}_{-0.93}$.}\\


\end{enumerate}

Future studies using data from surveys such as PRIMER \citep{primer} and the JWST Advanced Deep Extragalactic Survey (JADES), as well as near-infrared ground-based spectroscopy from the MOONS spectrograph on the VLT will provide higher SNR and larger samples of quiescent galaxies out to $z \simeq 2.5$. Combining these data with more sophisticated galaxy fitting methods (e.g. non-parametric SFHs) will enable a better understanding of quiescent galaxy properties and quenching mechanisms out to higher redshift.

\section*{Acknowledgements}
The authors thank the referee for useful comments which helped improve the quality of this manuscript. M. L. Hamadouche, R. Begley, and C. T. Donnan acknowledge the support of the UK Science and Technology Facilities Council. A. C. Carnall acknowledges the support of the Leverhulme Trust. 
F. Cullen and T. M. Stanton acknowledge support from a UKRI Frontier Research Guarantee Grant (PI Cullen; grant reference EP/X021025/1). Based on observations made with ESO Telescopes at the La Silla or Paranal Observatories under programme ID(s) 194.A-2003(E-Q) (The VANDELS ESO Public Spectroscopic Survey). Based on data products from observations made with ESO Telescopes at the La Silla Paranal Observatory under ESO programme ID 179.A-2005 and on data products produced by TERAPIX and the Cambridge Astronomy Survey Unit on behalf of the UltraVISTA consortium. This research made use of Astropy, a community-developed core Python package for Astronomy \citep{astropy:2013,astropy:2018}.

\section*{Data Availability}

The VANDELS survey is a European Southern Observatory Public Spectroscopic Survey. The full spectroscopic dataset, together with the complementary photometric information and derived quantities are available from \url{http://vandels.inaf.it}, as well as from the ESO archive \url{https://www.eso.org/qi/}. 
 
\color{black}
\bibliographystyle{mnras}
\bibliography{hamadouche2022}

\appendix

\bsp
\label{lastpage}
\end{document}